\documentclass[%
 reprint,
 amsmath,amssymb,
 aps,
floatfix,
]{revtex4-2}

\usepackage{graphicx}
\usepackage{dcolumn}
\usepackage{bm}
\usepackage{cleveref}

\begin{document}

\preprint{APS/123-QED}

\title{Non-Hermitian Dynamics Mimicked by a Hermitian Nonlinear System}

\author{Noah Flemens}
\email{nrf33@cornell.edu}
\author{Nicolas Swenson}
\author{Jeffrey Moses}
\email{moses@cornell.edu}
\affiliation{School of Applied and Engineering Physics, Cornell University, Ithaca, New York 14853, USA}%

\date{\today}

\begin{abstract}
We illustrate that a Hermitian nonlinear optical system consisting of hybridized parametric amplification and second harmonic generation mimics non-Hermitian evolution dynamics. Oscillation damping, evolution to a static steady state, and exceptional points arise from the use of second harmonic generation as an irreversible loss mechanism. The investigated system can be used to solve problems of inefficiency in parametric amplifier systems used widely in laser science and industrial applications. More generally, these findings suggest a new paradigm for the engineering of system dynamics where energy recovery and system sustainability are of importance.
\end{abstract}

\maketitle

A renewed interest in non-Hermitian systems over the past two decades has been fueled largely by the discovery that non-Hermitian Hamiltonians possessing PT-symmetry have real eigenspectra \cite{Bender:98}. The result of this effort has been a revolution in the ability to create and control evolution dynamics in a wide range of physical systems through the introduction of gain and loss, and because of the ease at which optical Hamiltonians can be experimentally implemented, many groundbreaking works have been in optical systems \cite{Ozdemir2019, El-Ganainy2018}.

The technique of introducing loss to a system to engineer the dynamics is of course old and ubiquitous. Shock absorbers in automobiles and resistance in electronic feedback loops are two of many familiar examples. In these systems, loss has been introduced to dampen unwanted oscillatory dynamics, providing convergence to a static steady state, a characteristic of non-Hermitian systems. However, avoiding the ill effects of loss (heating, wear, energy waste) has become an important engineering challenge for many technologies. Finding alternative ways of damping a system without the negative effects of loss -- for example, the use of regenerative braking in hybrid engines -- is key to producing sustainable and affordable technology and surpassing the state-of-the-art.

In this work, we report a prototypical nonlinear system that mimics the dynamics of a damped oscillator despite being a non-dissipative (fully conservative and closed), Hermitian system. This unusual behavior is brought about by the hybridization of multiple nonlinear three-wave mixing processes. Such processes are commonly found in fields such as electronics and optics, but arise more generally in systems with an anharmonic potential with nonzero cubic order.

Our investigation focuses on optical \textit{parametric} three-wave mixing processes, Hermitian processes in which the nonlinear electronic polarizability of a medium couples three light waves of different frequency or polarization, but no energy or momentum is exchanged with the medium. As \textit{Armstrong, et al.} pointed out in their seminal paper \cite{Armstrong:62}, this coupling can be thought of as thermal contact between electromagnetic modes. However, unlike in coupling to a thermal bath, a perpetual oscillation of power between the traveling waves occurs due to their maintained coherence. A spatiotemporal asynchronocity of these power oscillations -- due to the inhomogeneous transverse intensity profiles that are characteristic of real laser beams -- is highly detrimental to the efficiency of the integrated power exchange, as they result in nonuniform conversion. This is a main limiting factor in optical parametric amplification (OPA) \cite{Mironov_1982, Begishev:90, Moses:11}, a nonlinear process used to extend the frequency range of modern ultrafast laser systems. As a result, OPA, while an indispensable tool for ultrafast, strong-field, and relativistic optical science \cite{Cerullo2003, Mourou2006, Dubietis2006, Witte2012}, is considerably less efficient than laser amplification \cite{Mourou2006,Vaupel2013}. 
Recently, non-Hermitian variants of OPA were proposed -- named quasi-parametric amplification (QPA) or dissipative OPA -- in which the introduction of material losses to one of the three waves effectively damps out these power oscillations \cite{Ma:15, El-Ganainy:15, Zhong:16, Ma:17}. Despite a substantial dissipation of energy in the medium from one of the waves, the uniformity of spatiotemporal conversion between the other two waves was greatly improved, substantially boosting the amplifier efficiency.

Here we find that dynamics nearly identical to those of the non-Hermitian QPA system can be achieved in a Hermitian modified-OPA system, where a simultaneous parametric three-wave mixing process, second harmonic generation (SHG), is added to play the role of an irreversible loss mechanism. While hybridized nonlinear optical systems have been studied extensively \cite{Saltiel2005}, we find the system studied here -- which we call second harmonic amplification (SHA) -- possesses dynamics unlike those observed previously. The evolution dynamics of SHA exhibit damped oscillations and convergence to a static steady state. Unlike QPA, SHA is fully conservative, and all of the energy remains in coherent optical fields at the end of the device. To understand this, we introduce a Duffing oscillator model that unifies the description of OPA, QPA, and the newly introduced system of SHA under a generalized theoretical framework. We find that adding either loss or SHG to the so-called idler wave generated by OPA has the effect of adding damping to the Duffing oscillator model that describes OPA alone. Furthermore, an eigenvalue analysis uncovers that SHA mimics the non-Hermitian QPA system's damping characteristics and their relationship to broken and unbroken passive PT-symmetric phase. Finally, we employ a 3D spatiotemporal wave evolution analysis that captures the complex interaction of laser beams mixing in a bulk nonlinear medium during SHA. Net unidirectional energy flow over nearly the full spatiotemporal extent of the interacting waves is observed. Hence, we solve the problem of asynchronous conversion in OPA by an alternative method of introducing damping that does not involve loss. Such a method may solve longstanding problems of inefficiency in the frequency conversion of modern ultrafast laser systems. 

First, we present the OPA equations with SHG added as a loss mechanism. The quadratic nonlinear polarizability of a noncentrosymmetric medium mediates a one-to-two photon energy-conserving exchange, $\hbar\omega_l \leftrightarrow \hbar\omega_m + \hbar\omega_n$. A conservative power exchange between co-propagating electromagnetic fields at these frequencies occurs when they are far from any material resonances \cite{Armstrong:62}. Efficient power exchange requires a maintained coherence between propagating and nonlinear polarization fields. This occurs when the wave-vectors in the propagation medium obey the condition, $\mathbf{k_l}-\mathbf{k_m}-\mathbf{k_n} \equiv \mathbf{\Delta k} = \mathbf{0}$ (known as perfect phase matching) \cite{Giordmaine:62}.

The dynamics of SHA occur from a combination of phase-matched OPA (photon exchange: $\omega_p \leftrightarrow \omega_s + \omega_i$) and phase-matched SHG ($\omega_{2i} \leftrightarrow \omega_i + \omega_i$) with a shared frequency $\omega_i$. OPA employs three-wave mixing to transfer energy from a strong ``pump" field to a weak ``signal" field at a lower frequency, thus amplifying the signal while also producing an accompanying (and initially unpopulated) ``idler" field. Simultaneously, we employ SHG to upconvert the idler field to its initially unpopulated second harmonic (SH) at frequency $\omega_{2i} = 2\omega_i$. The propagation of monochromatic plane waves in the hybrid system can be modeled by four coupled evolution equations derived from Maxwell's equations assuming a nonzero macroscopic quadratic electric susceptibility:
\begin{align}
    d_{\zeta}u_p &= i u_s u_i e^{-i \Delta_{OPA} \zeta}\label{eq:ndsha-p} \\
    d_{\zeta}u_s &= i u_p u_i^* e^{i \Delta_{OPA} \zeta}\label{eq:ndsha-s} \\
    d_{\zeta}u_i &= i u_p u_s^* e^{i \Delta_{OPA} \zeta} + i 2\gamma_0 u_{2i} u_i^* e^{i \Delta_{SHG}\zeta}\label{eq:ndsha-i} \\
    d_{\zeta}u_{2i} &= i \gamma_0 u_i^2 e^{-i \Delta_{SHG} \zeta}. \label{eq:ndsha-2i}
\end{align}
\noindent The $u_j$ are nondimensional electric field amplitudes for $j \in \{p,s,i,2i\}$ where $\left| u_j \right|^2 = n_j$ is the fraction of photons in the $j$th field relative to the total number of initial photons, which we refer to as the fractional photon number of the $j$th field, $\Gamma_{OPA}$ and $\Gamma_{SHG}$ represent the drive strengths of the OPA and SHG processes, respectively, $\gamma_0=\Gamma_{SHG}/2\Gamma_{OPA}$ is the relative drive strength, $\zeta=\Gamma_{OPA} z$ is a nondimensionalized propagation coordinate, and $\Delta_{OPA} = (k_p-k_s-k_i)/\Gamma_{OPA}$ and $\Delta_{SHG} = (k_{2i}-2k_i)/\Gamma_{OPA}$ are the nondimensional wave-vector mismatches. (Definitions in terms of electric field complex amplitudes $A_j$, refractive indices, $\textrm{n}_j$, and nonlinear coefficient, $d_{\text{eff}}$: $k_j = \omega_j \textrm{n}_j / c$, $u_j=\sqrt{2 \textrm{n}_j \epsilon_0 c/\hbar \omega_j F_0}A_j$, $\Gamma_{OPA}=\sqrt{\hbar \omega_p \omega_s \omega_i d_{\text{eff}}^2 F_0/2 \textrm{n}_p \textrm{n}_s \textrm{n}_i \epsilon_0 c^3}$, and $\Gamma_{SHG}=\sqrt{\hbar \omega_i^2 \omega_{2i} d_{\text{eff}}^2 F_0/2 \textrm{n}_i^2 \textrm{n}_{2i} \epsilon_0 c^3}$. $F_0=\sum_j 2 \textrm{n}_j \epsilon_0 c\ |A_j(z=0)|^2/\hbar \omega_j$ is the initial photon flux. $d_{\rm{eff}}$ is proportional to the tensor element of the quadratic susceptibility for the specific field polarizations of the three mixing waves. For simplicity, we have assumed collinear waves where all frequencies are far below electronic resonances (i.e. negligible loss), and thus Kleinman symmetry implies $d_{\rm{eff}}$ is identical for the OPA and SHG processes \cite{Kleinman:62}.)

\begin{figure}[htbp]
	\centering
	    \includegraphics[width=3.4in]{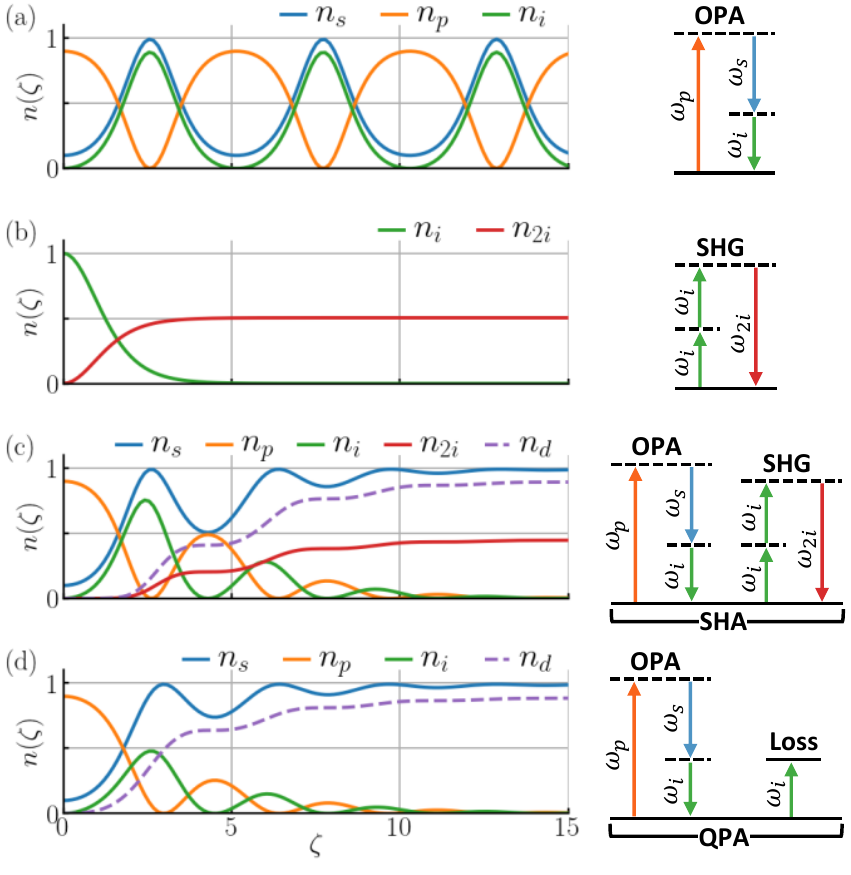}
        \caption{Fractional photon number exchange for different wave-mixing dynamics with $\gamma_0=0.35$ and $n_{p,0}=0.9$ under perfect phase matching conditions and corresponding photon mixing diagrams. (a) OPA: oscillatory exchange of photons between the pump and signal/idler fields. (b) SHG: unidirectional flow of idler photons to the idler SH field. (c) SHA: OPA with idler SHG displays the dynamics of a damped oscillator leading to a convergent flow of photons from the pump to the signal and idler SH fields. (d) QPA: similar dynamics to (c) when linear loss replaces the role of idler SHG.}
    \label{fig:dynamics}
\end{figure}
Figure \ref{fig:dynamics} compares the evolution dynamics of SHA with its isolated constituent parts, OPA and SHG. Numerical integration of \crefrange{eq:ndsha-p}{eq:ndsha-2i} can model the dynamics of phase-matched OPA when $\Delta_{OPA} = 0$ and $|\Delta_{SHG}| \to \infty$ (Fig. \ref{fig:dynamics}a) or phase-matched SHG when $\Delta_{SHG} = 0$ and $|\Delta_{OPA}| \to \infty$ (Fig. \ref{fig:dynamics}b). In Fig. \ref{fig:dynamics}a we use the initial conditions of OPA: $n_{p,0}\gg n_{s,0}$ and $n_{i,0}=0$ (and $n_{2i,0}=0$), and in Fig. \ref{fig:dynamics}b the initial conditions of SHG: $n_{i,0} = 1$ and $n_{2i,0}=0$ (and $n_{p,0}=n_{s,0}=0$). The OPA dynamics consist of undamped oscillations of power between pump and signal/idler fields and are described exactly by Jacobi elliptic functions for all values of $\Delta_{OPA}$ \cite{Armstrong:62,Milton:92,Moses:11}. In contrast, phase-matched SHG is a singular case of parametric three-wave mixing with non-oscillatory dynamics: as the idler field approaches zero, there are vanishing polarization fields at both frequencies (eqs. (\ref{eq:ndsha-i}), (\ref{eq:ndsha-2i})), thus preventing a reversal of the energy flow and slowing the exchange dynamics to a halt.

When $\Delta_{OPA} = \Delta_{SHG} = 0$, new four-wave dynamics emerge that are unlike those of either constituent wave-mixing process. Numerical integration of \crefrange{eq:ndsha-p}{eq:ndsha-2i} with the initial conditions for OPA result in evolution dynamics that evoke a damped oscillator beyond the initial depletion of the pump (Fig. \ref{fig:dynamics}c). A monotonic increase in the idler SH intensity steadily removes energy from the OPA system, which never returns to the initial state. Asymptotically, $n_p$ and $n_i$ reduce to 0 while $n_s$ and $n_{2i}$ approach 1 and $n_{p,0}/2$, respectively. 

To isolate the pump behavior, we make use of the two independent Manley-Rowe equations that describe conservation of fractional photon number (see \textit{Supplemental Material} for derivation \cite{Supplemental}),
\begin{align}
    1 &= n_p(\zeta) + n_s(\zeta) \label{eq:mr1} \\
    n_{p,0} &= n_p(\zeta) + n_i(\zeta) + n_d(\zeta), \label{eq:mr2}
\end{align}
\noindent where $n_d(\zeta)$ is the fraction of idler photons lost to the idler SH field (table \ref{tbl:duffing}). Under OPA initial conditions, and setting $\Delta_{OPA}=\Delta_{SHG}=0$, we differentiate eq. (\ref{eq:ndsha-p}) and combine with eqs. (\ref{eq:ndsha-s}), (\ref{eq:ndsha-i}), (\ref{eq:mr1}) and (\ref{eq:mr2}) to find:
\begin{equation}
    d_{\zeta}^2 u_p = -\left(1 + n_{p,0}-n_d(\zeta)\right)u_p + 2 u_p^3 - 2\gamma(\zeta) d_{\zeta} u_p \label{eq:duffing},\\
\end{equation}
\noindent where we have assumed $u_{p,0}$ and $u_{s,0}$ are positive real, from which it follows that $u_i = -u_i^*$ and $u_{2i}=-i\left| u_{2i} \right|$. 

Equation (\ref{eq:duffing}) is the force equation of a damped, undriven Duffing oscillator. The first term is a linear restoring force that decreases as the fraction of lost idler photons $n_d(\zeta)$ grows. The second term acts as a nonlinear softening of the restoring force. Under the constraints of OPA initial conditions and the Manley-Rowe equations, the sum of the first two terms of Eq. (\ref{eq:duffing}) must always be negative (i.e., the force is always restoring). The third term results in damping given by the coefficient $\gamma(\zeta) =\gamma_0 \sqrt{n_{2i}(\zeta)}$, which grows monotonically from zero as idler photons are unidirectionally displaced to the idler SH field. As $\gamma(\zeta)$ never switches sign, damping never switches to gain. Eq. (\ref{eq:duffing}) with eq. (\ref{eq:mr1}) reproduce the pump and signal behavior of Fig. \ref{fig:dynamics}c. We note that Eq. (\ref{eq:duffing}) with $n_d(\zeta)=\gamma(\zeta)=0$ (i.e., the case of conventional OPA, Fig. \ref{fig:dynamics}a) is the undamped, undriven cubic Duffing Equation, $d_{\zeta}^2 u_p = -(1+n_{p,0}) u_p + 2 u_p^3$, which is well known to have Jacobi elliptic function solutions \cite{Rand}. 
\begin{table}[htbp]
\centering
\begin{tabular}{ |c||c|c| }
    \hline
    Nonlinear Process & $n_d(\zeta)$ & $\gamma(\zeta)$\\
    \hline
    OPA & 0 & 0 \\
    \hline
    SHA & $2n_{2i}(\zeta)$ & $\gamma_0 \sqrt{n_{2i}(\zeta)}$ \\
    \hline
    QPA & $\frac{2\alpha}{\Gamma_{OPA}}\int^\zeta_0n_{i}(\zeta) d\zeta'$ & $\frac{\alpha}{2\Gamma_{OPA}}$\\
    \hline
\end{tabular}
\caption{Parameters of eqs. (\ref{eq:mr2}) to (\ref{eq:two-level}). $n_d(\zeta)$ is the fractional photon number displaced from the idler field by SHG (SHA) or loss (QPA) and $\gamma(\zeta)$ is a damping parameter.}
\label{tbl:duffing}
\end{table}

These dissipative oscillatory evolution dynamics reflect a departure from the behavior seen in prior works on simultaneous parametric three-wave-mixing processes, which have been studied widely in the contexts of effective cubic- (and higher) order phase modulation and wave mixing, multicolor solitons, and multiphoton generation \cite{Saltiel2005,Levenius:12, Jedrkiewicz2018, Kuo:20}. While many works experimentally observed simultaneously phase-matched SHG in OPA or optical parametric oscillator (OPO) systems, dynamics have been studied only through the first conversion peak (before SHA dynamics emerge), the range employed in practical devices.
A form of \crefrange{eq:ndsha-p}{eq:ndsha-2i} has been studied in the context of OPO \cite{Kartaloglu:03}, where the evolution dynamics are dominated by resonator feedback.
Other equations of motion for related hybridized systems have been derived and investigated, such as OPA with sum-frequency generation (SFG) \cite{Dikmelik1999} and cascaded SHG and SFG \cite{Petnikova:09, Sukhorukov2001}. These works identified analytical and numerical solutions in the form of \textit{periodic} multicomponent cnoidal functions, or irregular evolution dynamics. 

 The striking resemblance of SHA to a lossy system motivates an investigation for features usually associated with non-Hermitian systems. We treat the pump-idler subsystem as a linear two-level system with an adiabatically evolving coupling coefficient $u_s(\zeta)$ and monotonically increasing loss $\gamma(\zeta)$ given by the numerical solutions to eqs. (\ref{eq:ndsha-p}) - (\ref{eq:ndsha-2i}). To follow the pump and idler dynamics as damping occurs, we perform the gauge transformation $u_p = v_p e^{-\int_0^{\zeta}\gamma(\zeta')d\zeta'}$ and $u_i = v_i e^{-\int_0^{\zeta}\gamma(\zeta')d\zeta'}$ to reveal a simplified form \cite{Guo:09}. Substituting these coordinates into eqs. (\ref{eq:ndsha-p}) and (\ref{eq:ndsha-i}), we can write the coupled pump and idler equations in the form,
\begin{gather}
    -i\frac{d}{d\zeta}
    \begin{bmatrix}
        v_p \\
        v_i 
    \end{bmatrix}
    =
    \begin{bmatrix}
        - i\gamma(\zeta) & u_s(\zeta) \\
        u_s^*(\zeta) & i\gamma(\zeta) 
    \end{bmatrix}
    \begin{bmatrix}
        v_p \\
        v_i
    \end{bmatrix}.
    \label{eq:two-level}
\end{gather}
\noindent The operator of this system is non-Hermitian unless $\gamma(\zeta)=0$, which represents the case of conventional OPA. The eigenvalues are given by $\lambda = \pm \sqrt{n_s(\zeta) - \gamma(\zeta)^2}$. An exceptional point occurs when $n_s(\zeta)=\gamma(\zeta)^2$ or specifically, when $n_s(\zeta) = \gamma_0^2 n_{2i}(\zeta)$. When $n_s(\zeta) > \gamma(\zeta)^2$, the operator is PT-symmetric with purely real eigenvalues. When $n_s(\zeta)<\gamma(\zeta)^2$, PT-symmetry is broken, resulting in purely imaginary eigenvalues. Consequently, for values of $\gamma(\infty)<1$, PT-symmetry is never broken and the pump and idler fields oscillate forever (Fig. \ref{fig:PTSymmetry}a, b) (see \textit{Supplemental Material} for proof \cite{Supplemental}). When $\gamma(\infty)>1$, PT-symmetry is eventually broken, leading to non-oscillatory, exponential growth of the pump and idler fields in the gauge transformed frame (Fig. \ref{fig:PTSymmetry}c, d). 

\begin{figure}[htbp]
	\centering
	    \includegraphics[width=3.41in]{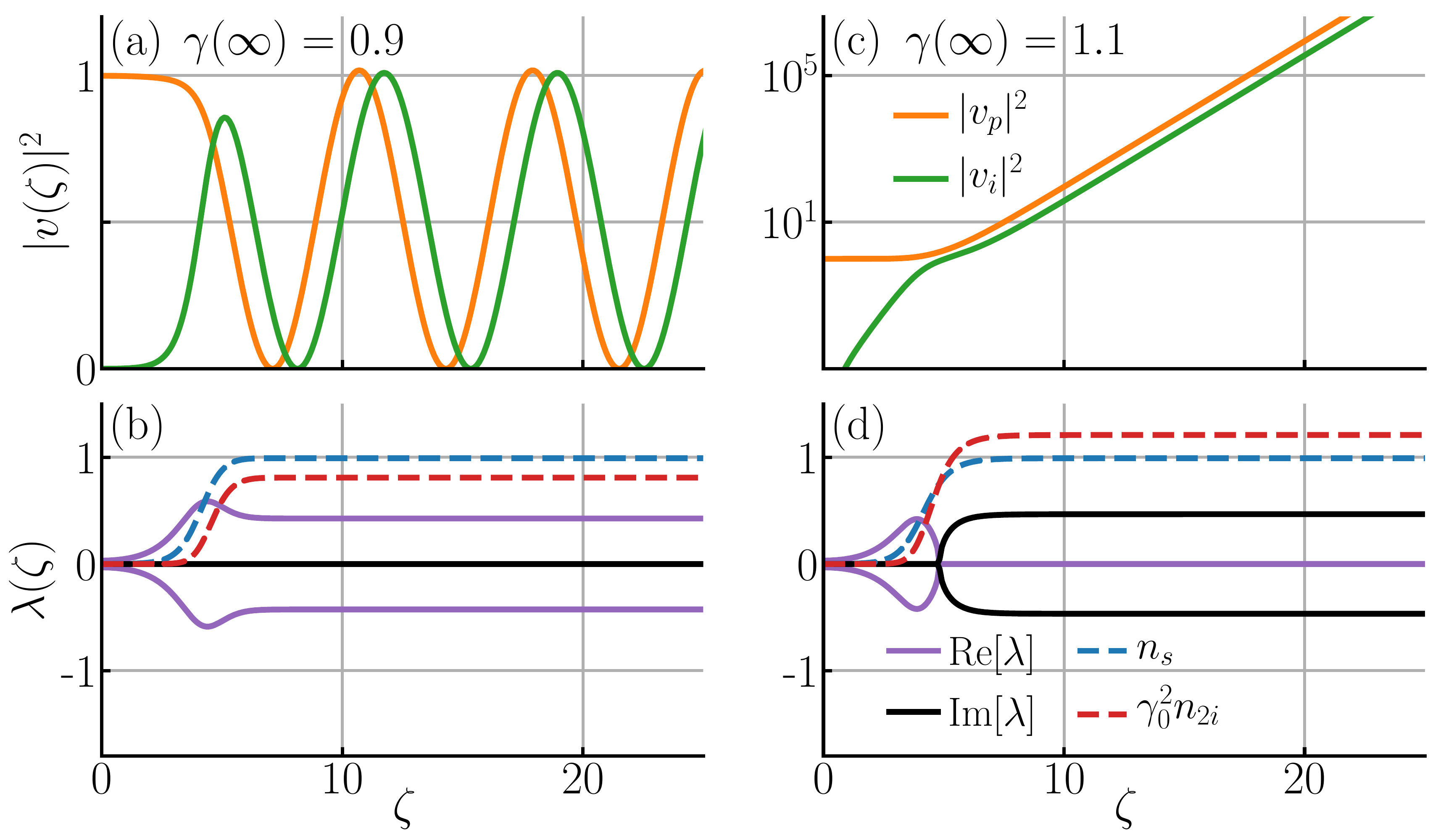}
        \caption{Pump and idler dynamics in the gauge transformed frame showing for $\gamma(\infty)<1$ (a,b), power oscillations and purely real eigenvalues for all $\zeta$, and for $\gamma(\infty)>1$ (c,d), exponential growth and a transition from purely real to purely imaginary eigenvalues at the exceptional point at $n_s=\gamma_0^2 n_{2i}$.}
    \label{fig:PTSymmetry}
\end{figure}

\begin{figure}[htbp]
	\centering
	    \includegraphics[width=3.41in]{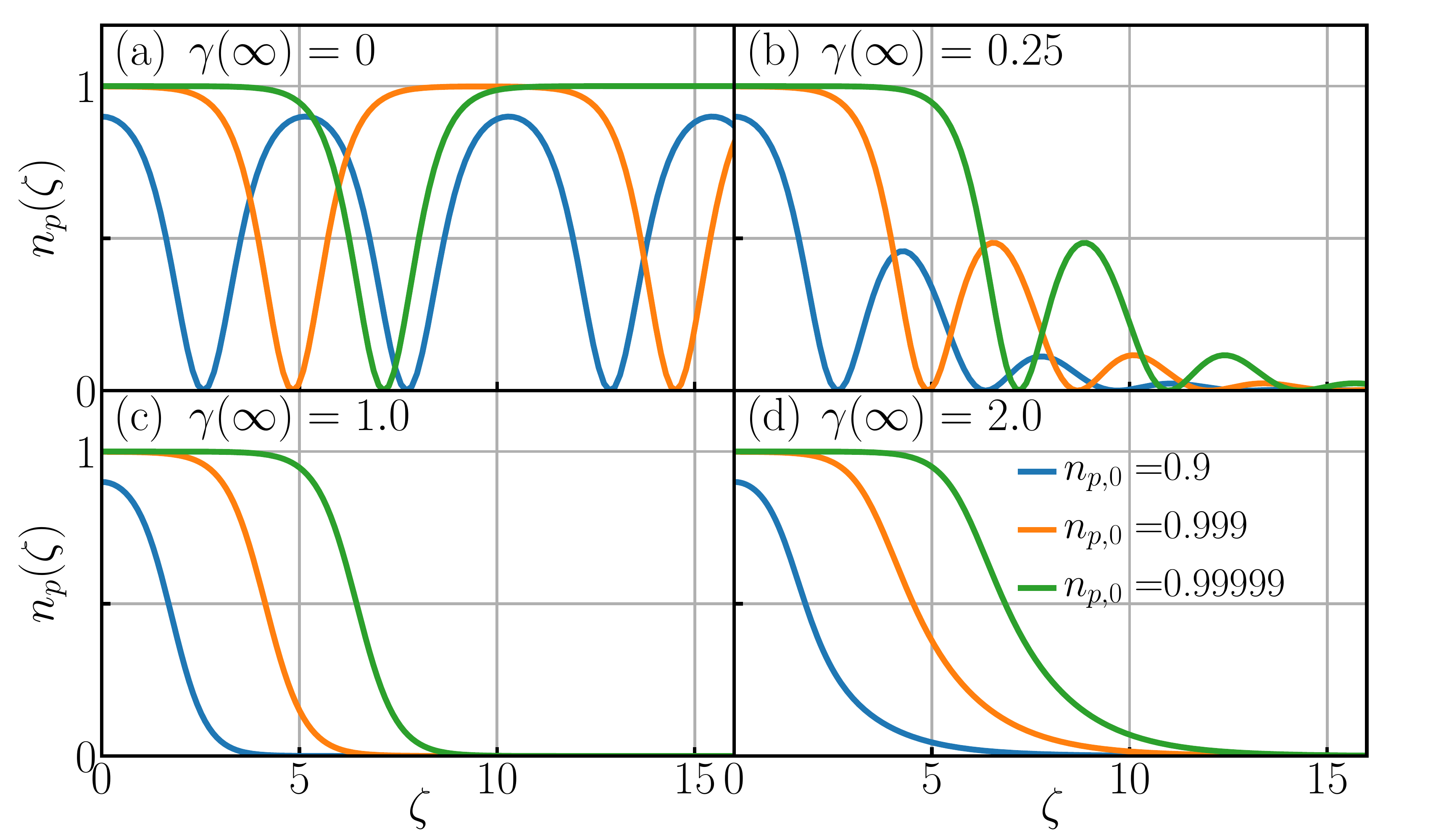}
        \caption{Pump fractional photon number evolution under various damping conditions and values of initial pump photon number: (a) conventional OPA with no damping, (b) underdamped SHA (real eigenvalues), (c) critically damped SHA (asymptotic approach to the exceptional point), and (d) overdamped SHA (transition from real to imaginary eigenvalues).}
    \label{fig:damping}
\end{figure}

We find these regimes correspond to the underdamped and overdamped regimes of an oscillator, with critical damping occurring when $\gamma(\infty)=1$ where the exceptional point is reached at $\zeta \rightarrow \infty$. Fig. \ref{fig:damping} shows a numerical solution of \crefrange{eq:ndsha-p}{eq:ndsha-2i} for $\gamma(\infty)=\gamma_0 \sqrt{n_{2i}(\infty)}=\sqrt{\textrm{n}_p \textrm{n}_s \omega_i \omega_{2i} n_{p,0}/8 \textrm{n}_i \textrm{n}_{2i} \omega_p  \omega_s} =$ 0, 0.25, 1, and 2, and clearly depicts the regimes of a damped oscillator: undamped, underdamped, critically damped, and overdamped, respectively. For each value of $\gamma(\infty)$, the dynamics of $n_p$ are shown for three values of $n_{p,0}$ (0.9, 0.999, 0.99999), roughly corresponding to 10, 30, and 50 dB signal gain as $\zeta \rightarrow \infty$. The effect of a larger $n_{p,0}$ (higher gain) is primarily a delay of the dynamics. When damped, the universal convergence to zero amplitude is key to enhancing the efficiency of OPA for real laser beams, as will be discussed below.

The appearance of non-Hermitian characteristics in a fully conservative, closed system is remarkable, and begs comparison to the closest system with actual loss, i.e. QPA, where irreversible removal of idler photons through SHG is replaced with linear absorption by means of a dopant material \cite{Ma:15,Ma:17}. This is modeled by replacement of \cref{eq:ndsha-i} with 
\begin{equation}
    d_{\zeta}u_i = i u_p u_s^* e^{i \Delta_{OPA} \zeta} -\frac{\alpha}{\Gamma_{OPA}} u_i\label{eq:ndqpa-i},
\end{equation}
\noindent where the linear loss coefficient, $\alpha$, is assumed real and positive. Numerical integration of eqs. (\ref{eq:ndsha-p}), (\ref{eq:ndsha-s}) and (\ref{eq:ndqpa-i}) (Fig. \ref{fig:dynamics}d) illustrates evolution dynamics remarkably similar to those of SHA. These equations can be cast in the Duffing oscillator form, eq. (\ref{eq:duffing}), by replacing $n_d(\zeta)$ and $\gamma(\zeta)$ with the values listed for QPA in table \ref{tbl:duffing}. Furthermore, using the same gauge transformation as above, eqs. (\ref{eq:ndsha-p}) and (\ref{eq:ndqpa-i}) can be cast in the form of eq. (\ref{eq:two-level}). An exceptional point analysis reveals damping regimes separated by the critical damping condition $n_s(\infty)=\alpha/2\Gamma_{OPA}$. In this light, the fully Hermitian SHA system closely mimics the non-Hermitian behavior of QPA, with the subtle difference of monotonically evolving, rather than constant, diagonal terms in the sub-system operator. Clearly, the non-Hermitian pump-idler subsystem is agnostic to how energy is being displaced from the idler field. Of course, the distinctly different physical damping mechanisms pose different practical consequences. In QPA, idler photons are dissipated by material absorption and lost as heat, while in SHA, a nonlinear polarizability irreversibly converts them to a coherent copropagating wave. At the end of the medium, this wave can be separated from the amplified signal by a beamsplitter, allowing complete removal and possible reuse of its energy.

We now investigate the four-wave evolution dynamics of SHA using a propagation model that captures the spatiotemporal dynamics of picosecond laser beams in a realistic setting (see \textit{Supplemental Material} for further details \cite{Supplemental}). Our example addresses the OPA efficiency problem: the inhomogeneous spatial and temporal intensity profiles of real laser beams make uniform pump depletion unfeasible in OPA, since sensitivity of the field power exchange cycle period to initial pump and signal photon number makes the length for full conversion vary with transverse coordinate. Historically, flattop or conformal laser profile shaping has been proposed, and in a few cases achieved, to synchronize power oscillations across the spatiotemporal coordinate and improve conversion efficiency \cite{Begishev:90, Moses:11, Waxer:03, Bagnoud:05, Fulop:07, Cao:18}. However, as shaping is usually impractical, low pump-to-signal energy conversion efficiencies of a few to twenty percent plague most OPA systems \cite{Vaupel2013}. The damped power oscillations of QPA were proposed as a solution \cite{Ma:15}, but the need for engineered materials and the accompanying strong absorption of laser power has so far limited its use.

As found generally for hybridized parametric processes \cite{Saltiel2005}, we find that commonly used nonlinear materials can support simultaneous birefringent phase matching of SHG and OPA over a broad range of pump wavelengths, enabling the hybridized process of SHA. We modeled the SHA interaction in CdSiP$_2$ (CSP), a birefringent material relevant to mid-infrared applications of OPA \cite{Sanchez:14,Liang:17}. At a crystal orientation of $\theta=44.8^\circ$, SHA is phase matched for the Ho:YLF laser wavelength 2.05 $\mu$m (pump) and a 3.0-$\mu$m signal. These wavelengths correspond to a 6.5-$\mu$m idler and 3.25-$\mu$m idler SH. To reflect a realistic laboratory laser beam interaction, the initial pump and signal spatiotemporal profiles were 1st-order Gaussian in space and time, with 1 mm and 2 mm beam radii ($1/e^2$ half width) respectively and 1 ps pulse durations (in FWHM). The initial pump energy was 0.87 mJ, resulting in a peak intensity (50 GW/cm$^2$) below the damage threshold of CSP. The initial signal energy was chosen to be 2 nJ. These parameters correspond to $\gamma_0=0.27$.

\begin{figure}[htbp]
    \centering
        \includegraphics[width=3.41in]{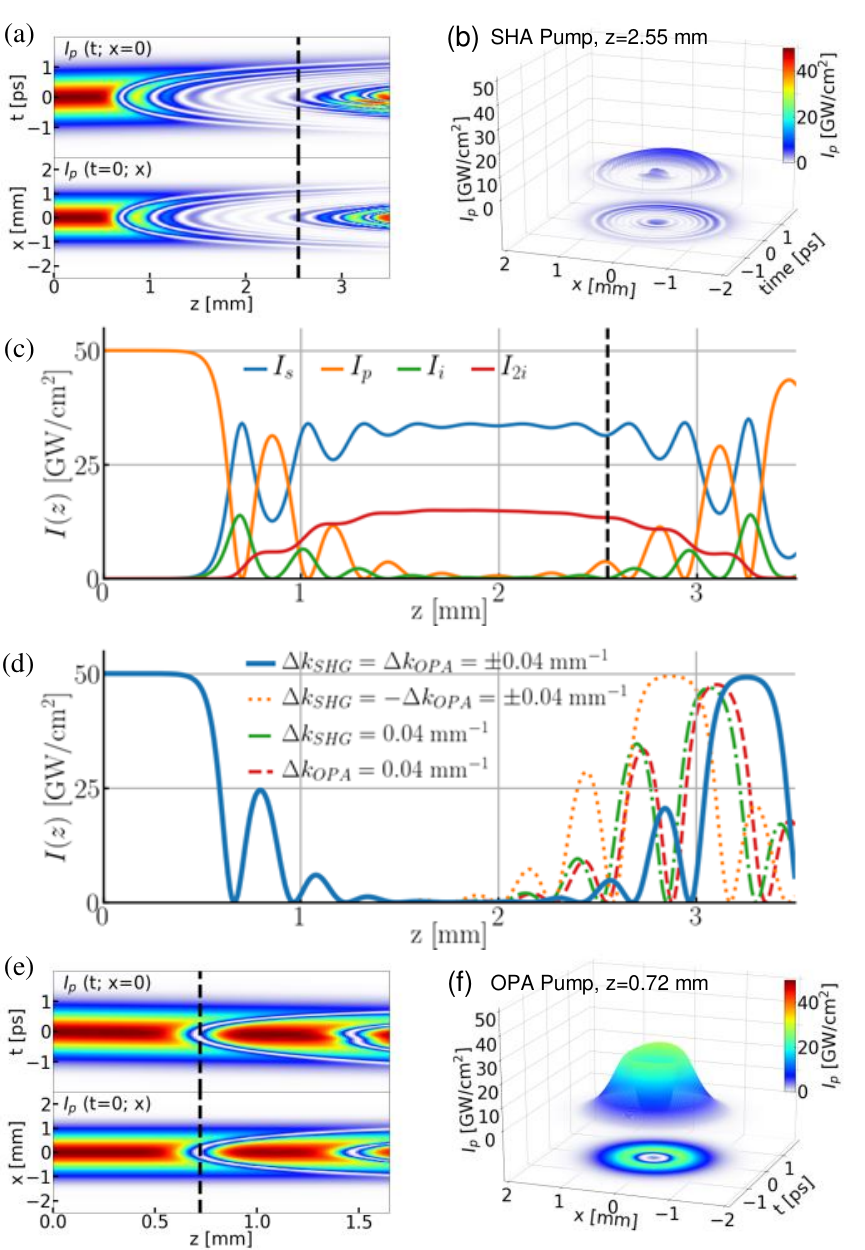}
        \caption{(a) SHA spatiotemporal pump intensity depletion dynamics. Dashed line: length for optimum conversion efficiency. (b) Residual pump intensity profile at the optimum conversion length resulting in 80.0\% pump energy depletion and 54.6\% energy conversion to the signal. (c) Intensity evolution of the four fields at $t=x=0$ showing eventual back-conversion. (d) 1D simulation of pump intensity for nonzero wave-vector mismatch. (e) Pump depletion dynamics as in (a) but for conventional OPA. Dashed line: length for optimum conversion efficiency. (f) Residual pump intensity profile at the optimum conversion length resulting in only 13\% pump energy depletion and 9\% energy conversion to the signal.}
        \label{fig:cspresults}
\end{figure}

Figure \ref{fig:cspresults}a shows the spatiotemporal conversion dynamics of underdamped SHA, which proceed as in conventional OPA until the first half conversion cycle, after which subsequent conversion cycles are rapidly damped. For the majority of the interaction length we observe a gradual approach toward full spatiotemporal pump depletion. The spatiotemporal asynchronicity of conversion cycles no longer prevents uniform depletion, as there is a common length at which the cycles are well damped. However, we observe a reversal of flow not predicted by Fig. \ref{fig:dynamics}c that returns energy to the pump field, resulting in a pump depletion maximum at $z = $ 2.55 mm. Cutting the medium at this length results in 80.0\% energy depletion of the pump (Fig. \ref{fig:cspresults}b) with 54.6\% energy conversion to the signal and the remaining energy split between the idler SH (22.4\%) and idler (3.0\%).

Since pulsed laser beams possess a distribution of wavevectors and frequencies, some phase mismatch is inherent in any real application. Fig. \ref{fig:cspresults}c shows the intensity dynamics at the center of each pulse ($(x,t)=(0,0)$). At $z \gtrsim$ 2 mm, energy can be observed returning from the idler SH to the idler field and, subsequently, from the signal and idler to the pump field. Numerical integration of \crefrange{eq:ndsha-p}{eq:ndsha-2i} including $\pm$0.04 mm$^{-1}$ of wave-vector mismatch (Fig. \ref{fig:cspresults}d) suggests the back-conversion dynamics seen in Figs. \ref{fig:cspresults}a, c to be the result of phase mismatch. 

The practical benefits of SHA become clear upon repeating the same analysis for conventional OPA (Fig. \ref{fig:cspresults}e, implemented numerically by setting $\Delta k_{SHG} \rightarrow \infty$). Undamped conversion cycles typical of conventional OPA are observed. These are strongly asynchronous across the spatiotemporal profile, giving rise to poor integrated energy transfer. A device length of 0.7 mm maximizes conversion efficiency before significant back conversion of the signal occurs with only 13\% pump energy depletion and 9\% energy conversion to the signal (Fig. \ref{fig:cspresults}f). This is 6 times smaller than the efficiency achievable by SHA.

We envision many potential applications enabled by the nearly full spatiotemporal pump depletion afforded by SHA, which improves the efficiency of both signal amplification and conversion to the idler SH. To extend the range of systems that fulfill the phase-matching condition, quasi-phase matching may be used \cite{Chou:99, Liu:02, Lifshitz:05, Saltiel2005}. It may even become possible to surpass the quantum defect limit of OPA efficiency (i.e., an efficiency $\leq \omega_s/\omega_p$), as the substantial power in the idler SH field could be reused in a second SHA stage to further amplify the signal.

In summary, we have demonstrated a closed conservative system in nonlinear optics, SHA, that possesses common behaviors of non-Hermitian systems: distinct phases of unbroken and broken PT symmetry separated by an exceptional point in its two-mode subsystem, characterized by damped oscillations and exponential decay, respectively. This is brought about by a nonlinear wave mixing process, phase-matched SHG, that mimics the irreversible energy flow of thermal contact with a heat bath. Using SHG as a loss mechanism on a second three-wave mixing system characterized by the usual perpetual oscillation between modes, OPA, allowed engineering of the evolution to closely match that of OPA with loss (QPA). We have further shown that a Duffing oscillator model unifies the description of OPA with and without the added characteristics of idler loss (QPA) or idler SHG (SHA). As verified by a 3D spatiotemporal analysis of SHA for a realistic laboratory implementation, we found that the longstanding technical hurdle of poor OPA efficiency resulting from spatiotemporally nonuniform conversion can be averted. The damped oscillatory behavior of SHA allows this while efficiently generating a reusable second coherent wave and avoiding the practical costs of implementing and incurring system losses.

The observed non-Hermitian dynamics of SHA are distinct from those seen previously in any fully parametric nonlinear optical system. This motivates a further search for non-Hermitian behavior in parametric processes hybridized with SHG (or higher-order harmonic generation of the form $\omega_n = n\omega_0$), in optics and in other physical systems that allow wave mixing through an anharmonic potential. Generally, the unusual phenomenon of oscillation damping in a closed, conservative system may find use in the engineering of system dynamics where energy recovery and system sustainability are of importance. 

Underlying data are available at Ref. \cite{Repository}. This work was supported equally by the Cornell Center for Materials Research with funding from the NSF MRSEC program (DMR-1719875), which provided initial support, and by the NSF under grant no. ECCS-1944653.


\providecommand{\noopsort}[1]{}\providecommand{\singleletter}[1]{#1}%

\end{document}


\title{Non-Hermitian Dynamics Mimicked by a Hermitian Nonlinear System: Supplementary Material}

\author{Noah Flemens}
\author{Nicolas Swenson}
\author{Jeffrey Moses}
\affiliation{School of Applied and Engineering Physics, Cornell University, Ithaca, New York, 14853, USA}
\affiliation{Corresponding authors: nrf33@cornell.edu, moses@cornell.edu}



\begin{abstract}
This document provides supplementary information to “Non-Hermitian Dynamics Mimicked by a Hermitian Nonlinear System,” 
\end{abstract}

\maketitle
\subsection{Derivation of Manley-Rowe equations}

The Manley-Rowe equations express conservation of fractional photon number and can be derived by taking the derivatives of the fractional photon number of each field with respect to $\zeta$ and substituting the coupled wave equation for each respective field (we drop the explicit zeta dependence of the fields for clarity):

\begin{align*}
    d_{\zeta}n_p & = d_{\zeta}|u_p|^2\\
                 & = u_p^* d_{\zeta}u_p + u_p d_{\zeta}u_p^*\\
                 & = i u_p^* u_s u_i e^{-i\Delta_{OPA}\zeta} - i u_p u_s^* u_i^* e^{i\Delta_{OPA}\zeta}\\
    d_{\zeta}n_s & = d_{\zeta}|u_s|^2\\
                 & = u_s^* d_{\zeta}u_s + u_s d_{\zeta}u_s^*\\
                 & = i u_p u_s^* u_i^* e^{i\Delta_{OPA}\zeta} - i u_p^* u_s u_i e^{-i\Delta_{OPA}\zeta}\\
    d_{\zeta}n_i & = d_{\zeta}|u_i|^2\\
                 & = u_i^* d_{\zeta}u_i + u_i d_{\zeta}u_i^*\\
                 & = i u_p u_s^* u_i^* e^{i\Delta_{OPA}\zeta} - i u_p^* u_s u_i e^{-i\Delta_{OPA}\zeta}\\ 
                 & + i 2\gamma_0 u_{2i} (u_i^*)^2 e^{i\Delta_{SHG}\zeta} - i 2\gamma_0 u_{2i}^* u_i^2 e^{-i\Delta_{SHG}\zeta}\\
    d_{\zeta}n_{2i} & = d_{\zeta}|u_{2i}|^2\\
                 & = u_{2i}^* d_{\zeta}u_{2i} + u_{2i} d_{\zeta}u_{2i}^*\\
                 & = i \gamma_0 u_{2i}^* u_i^2 e^{-i\Delta_{SHG}\zeta} - i \gamma_0 u_{2i} (u_i^*)^2 e^{i\Delta_{SHG}\zeta}
\end{align*}

\noindent The Manley-Rowe equations are then found by noting:
\begin{align*}
    && d_{\zeta}n_p &= -d_{\zeta}n_s\\
    \implies && d_{\zeta}n_p + d_{\zeta}n_s &= 0\\
    \implies && n_p + n_s &= C_1
\end{align*}

\noindent and

\begin{align*}
    && d_{\zeta}n_i &= -d_{\zeta}n_p - 2d_{\zeta}n_{2i}\\
    \implies && d_{\zeta}n_i + d_{\zeta}n_p + 2d_{\zeta}n_{2i} &= 0\\
    \implies && n_p + n_i + 2n_{2i} &= C_2
\end{align*}

\noindent where $C_1$ and $C_2$ are constants. Under the initial conditions of OPA, $n_{p,0}>n_{s,0}>0$ and $n_{i,0}=n_{2i,0}=0$. Hence,

\begin{align*}
    1 &= n_p + n_s\\
    n_{p,0} &= n_p + n_i + 2n_{2i}
\end{align*}

\noindent where we have used the fact that $\Sigma n_{j,0} = 1$ for $j \in \{p, s, i, 2i \}$ for the first equation.

For the case of QPA, the idler equation becomes:
\begin{align*}
    d_{\zeta}n_i & = d_{\zeta}|u_i|^2\\
                 & = u_i^* d_{\zeta}u_i + u_i d_{\zeta}u_i^*\\
                 & = i u_p u_s^* u_i^* e^{i\Delta_{OPA}\zeta} - i u_p^* u_s u_i e^{-i\Delta_{OPA}\zeta}\\ 
                 & -\frac{2\alpha}{\Gamma_{OPA}} |u_i|^2\\
\end{align*}

\noindent While the Manley-Rowe equation relating the pump and signal remain unchanged, the equation for the pump and idler is found from noting:
\begin{align*}
    && d_{\zeta}n_p &= -d_{\zeta}n_i - \frac{2\alpha}{\Gamma_{OPA}} n_i\\
    \implies && d_{\zeta}n_p + d_{\zeta}n_i + \frac{2\alpha}{\Gamma_{OPA}} n_i &= 0\\
    \implies && n_p + n_i + \frac{2\alpha}{\Gamma_{OPA}}\int^\zeta_0 n_{i} d\zeta' &= C_3
\end{align*}

\noindent where $C_3$ is a constant. Again, from the initial conditions of OPA, we find:
\begin{align*}
    n_{p,0} &= n_p + n_i + \frac{2\alpha}{\Gamma_{OPA}}\int^\zeta_0n_{i} d\zeta'
\end{align*}

\subsection{Proof: $\gamma(\infty)<1 \rightarrow Im \{ \lambda(\zeta) \}=0$ $\forall$ $\zeta$}

From the main text, the eigenvalues for the pump-idler subsystem of SHA are given by $\lambda(\zeta)=\pm \sqrt{n_s(\zeta) - \gamma_0^2 n_{2i}(\zeta)}$. Clearly, $Im \{ \lambda(\zeta) \}=0$ when $n_s(\zeta) > \gamma_0^2 n_{2i}(\zeta)$ for all values of $\zeta$. Starting with the assumption that $\gamma(\infty)<1$,  then:
\begin{align*}
    && \gamma(\infty) &< 1\\
    \implies && \gamma_0^2 n_{2i}(\infty)& < 1\\
    \implies && \gamma_0^2 &< \frac{2}{n_{p,0}}\\
    \implies && \gamma_0^2 n_{2i}(\zeta) &< \frac{2}{n_{p,0}} n_{2i}(\zeta) 
\end{align*}
\noindent where we have used the fact that $n_{2i}(\zeta)$ monotonically increases to $n_{p,0}/2$ as $\zeta \rightarrow \infty$. 

Now we wish to show $n_s(\zeta) \geq \frac{2}{n_{p,0}} n_{2i}(\zeta) > \gamma_0^2 n_{2i}(\zeta)$. By way of contradiction, assume that $n_{p,0} n_s(\zeta) < 2 n_{2i}(\zeta)$, then the Manley-Rowe relations state:
\begin{align*}
    && n_p(\zeta) + n_i(\zeta) + 2n_{2i}(\zeta) &= n_{p,0}\\
    \implies && n_p(\zeta) + n_i(\zeta) + n_{p,0}n_s(\zeta) &< n_{p,0}\\
    \implies && n_p(\zeta) + n_i(\zeta) &< n_{p,0} (1-n_s(\zeta))\\
    \implies && n_i(\zeta) &< n_{p,0} n_p(\zeta) - n_p(\zeta)\\
    \implies && n_i(\zeta) &< -n_{s,0} n_p(\zeta)\\
    \implies && n_i(\zeta) &< 0
\end{align*}
\noindent where we have used $n_p(\zeta)+n_s(\zeta)=1$ and $n_{s,0} n_p(\zeta) > 0$. This is a contradiction since $n_i$ must be greater than $0$, thus: 
\begin{align*}
&& n_{p,0} n_s(\zeta) &\geq 2 n_{2i}(\zeta)\\
\implies && n_s(\zeta) &\geq \frac{2}{n_{p,0}} n_{2i}(\zeta)\\
\implies && n_s(\zeta) &> \gamma_0^2 n_{2i}(\zeta)\\
\implies && Im\{ \lambda(\zeta) \} &= 0 \hspace{0.5cm} \forall \zeta
\end{align*}

\noindent Therefore, the eigenvalues are purely real for all $\zeta$ when $\gamma(\infty)<1$ and the pump and idler modes will oscillate forever.

\subsection{Evaluation of diffraction and beam walk-off in SHA}

We performed an initial analysis of beam propagation effects in order to find the smallest beam size where diffraction and Poynting vector walk-off are negligible for the SHA device considered in our study. Monochromatic fields at frequencies $\omega_{j,0}$ for signal, pump, idler, and idler second harmonic (SH) waves with 1D spatial Gaussian beam profiles were propagated using the four coupled pulse propagation equations for OPA and idler SHG (shown in the spatial Fourier domain) with diffraction and Poynting vector walk-off:
\begin{align}
    d_{z}E_s(k_x) = i& \frac{\omega_{s,0} d_{\text{eff}}}{n_s c} \mathcal{F} \{E_p(x) E_i^*(x)\} - i k_s(\omega_{s,0})E_s(k_x) \nonumber \\
    +& i \frac{k_x^2}{2k_s(\omega_{s,0})} E_s \label{eq:space-s} \\
    d_{z}E_p(k_x) = i& \frac{\omega_{p,0} d_{\text{eff}}}{n_p c} \mathcal{F} \{E_s(x) E_i(x)\} - i k_p(\omega_{p,0})E_p(k_x) \nonumber \\
    +& i \rho_{p} k_x E_{p}(k_x) + i \frac{k_x^2}{2k_p(\omega_{p,0})} E_p(k_x) \label{eq:space-p} \\
    d_{z}E_i(k_x) = i& \frac{\omega_{i,0} d_{\text{eff}}}{n_i c}\mathcal{F} \left\{E_p(x) E_s^*(x) + E_{2i}(x) E_i^*(x)\right\} \nonumber \\ 
    -& i k_i(\omega_{i,0})E_i(k_x) + i \frac{k_x^2}{2k_i(\omega_{i,0})} E_i(k_x) \label{eq:space-i} \\
    d_{z}E_{2i}(k_x) = i& \frac{\omega_{2i,0} d_{\text{eff}}}{2n_{2i} c} \mathcal{F} \{E_i^2(x)\} - i k_{2i}(\omega_{2i,0})E_{2i}(k_x) \nonumber \\
    +& i \rho_{2i} k_x E_{2i}(k_x) + i \frac{k_x^2}{2k_{2i}(\omega_{2i,0})} E_{2i}(k_x) \label{eq:space-2i}
\end{align}
\noindent where  $E_j(k_x)=A_j (k_x)e^{i k_j(\omega_{j,0})z}$, and $A_j$, $k_j$, and $n_j$ are the signal, pump, idler, and idler SH electric field amplitudes, wave vectors in the nonlinear media, and indices of refraction, respectively, and where $k_x$ and $z$ are the transverse spatial and propagation coordinates. The effective quadratic nonlinear coefficient is given by $d_{eff}$ and $c$ is the speed of light. Terms quadratic in $k_x$ represent diffraction in the paraxial regime. Terms linear in $k_x$ represent Poynting vector walk-off, which in CSP is present for pump and idler SH but not signal and idler, as both OPA and SHG processes are Type-I ($o + o = e$). The walk-off angles for the pump and idler SH are given by $\rho_p$ and $\rho_{2i}$, respectively.

\begin{figure}[htbp]
    \centering
        \includegraphics[width=3.46in]{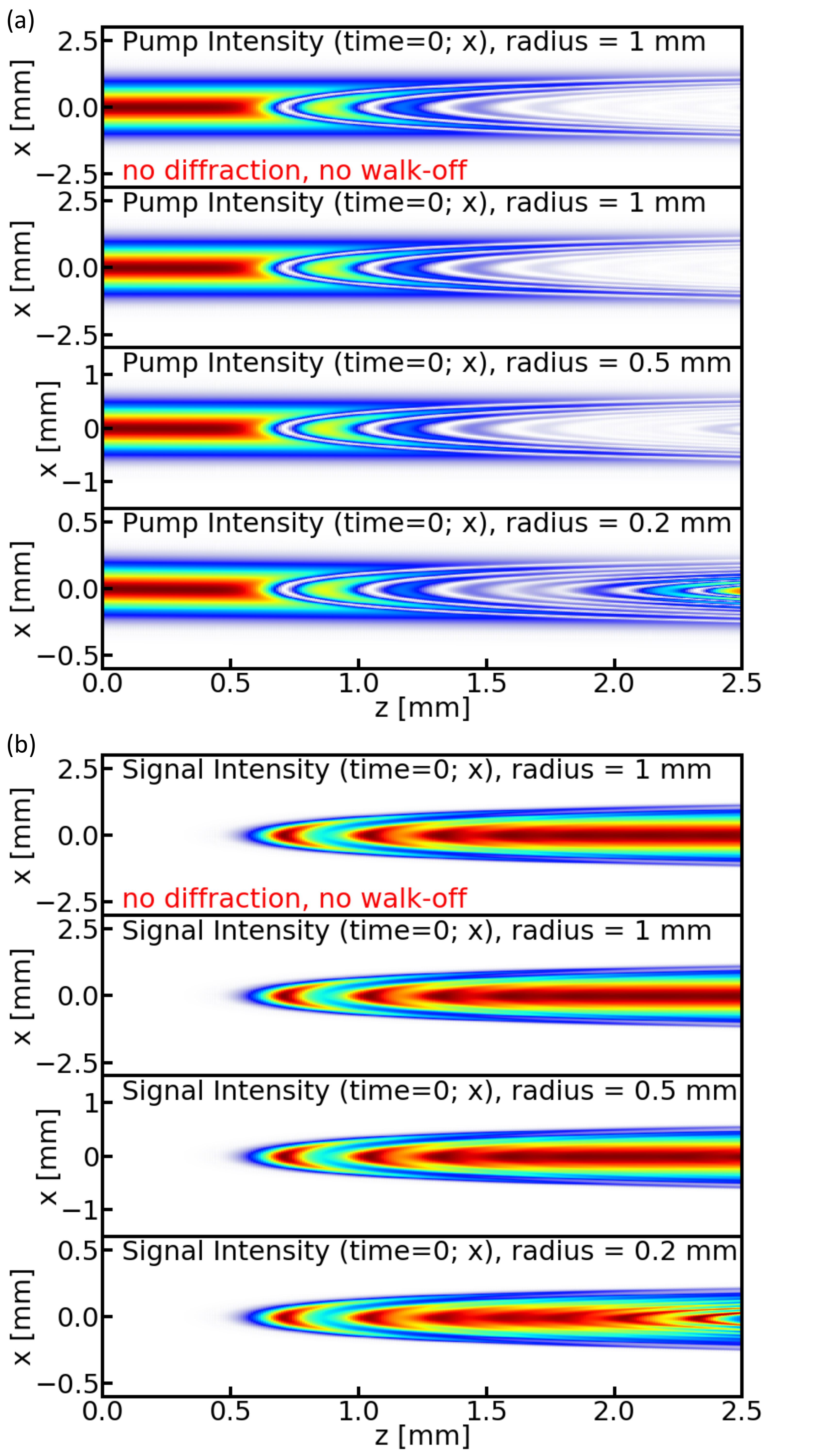}
        \caption{Spatial beam propagation dynamics for Type-I ($ooe$) phase-matched OPA and Idler SHG in CSP at time = 0 for (a) pump beams of radius 1, 0.5, and 0.2 mm (1/$e^2$) and (b) the corresponding signal. The first panels of (a) and (b) show the 1 mm beam radius case where all spatial propagation effects are neglected from the simulation. Simulation parameters correspond to the device parameters in the main text.}
    \label{fig:spatialresults}
\end{figure}

Fig. \ref{fig:spatialresults} shows the results for the CSP device, where both diffraction and spatial walk-off are of concern. $\rho_p$ and $\rho_{2i}$ for the 1.03 $\mu$m pump and 3.25 $\mu$m idler SH in this case are 16.64 $\mu$rad and 16.82 $\mu$rad, respectively. For the 2.55 mm optimal crystal length used in the simulation, this corresponds to ~40 $\mu$m of walk-off for the pump. Fig. 1 shows the spatial simulation results of solving Eqs. (S1)-(S4) for three beam sizes, $1/e^2$ beam radii 1, 0.5, and 0.2 mm. For a 1 mm beam radius (the pump beam radius used in the CSP device example in the main text), we find negligible contribution from diffraction and walk-off, resulting in dynamics identical to the simulations where these effects are not included (top two panels). Decreasing to a 0.5 mm beam radius, a very slight deviation appears in the last 10\% of the crystal length, with minor back-conversion setting in just as the crystal terminates. Decreasing even further to a 0.2 mm beam radius, there is no longer a region within the crystal where conversion cycles are fully damped out. These results show that ignoring spatial effects is an excellent approximation for a 1.0 mm pump beam radius and marginal at a 0.5 mm beam radius. At 0.2 mm radius, diffraction and spatial walk-off have a strong effect on the evolution.

This analysis illustrates that spatial effects can disturb the SHA dynamics, preventing uniform spatiotemporal amplification. However, the use of suitably large beams circumvents the problem.

\subsection{Numerical model used for full spatiotemporal analysis of SHA propagation}

For the CSP device simulated in the main text, where diffraction and spatial walk-off were found to have negligible effect as a result of the large beam size, full spatiotemporal evolution (two transverse spatial dimensions plus one temporal dimension) along the propagation axis could be solved independently for each transverse spatial coordinate ($x,y$), with a temporal grid to capture temporal propagation effects. Using a Fourier split-step method, we solved the following four coupled pulse propagation equations for OPA and idler SHG (shown in the frequency domain), accounting for the exact frequency-dependent dispersion, $k_j(\omega)$, given by published Sellmeier equations:
\begin{align}
    d_{z}E_s(\omega) = i& \frac{\omega_{s,0} d_{\text{eff}}}{n_s c} \mathcal{F} \{E_p(t) E_i^*(t)\} - i k_s(\omega)E_s(\omega) \label{eq:time-s} \\
    d_{z}E_p(\omega) = i& \frac{\omega_{p,0} d_{\text{eff}}}{n_p c} \mathcal{F} \{E_s(t) E_i(t)\} - i k_p(\omega)E_p(\omega) \label{eq:time-p} \\
    d_{z}E_i(\omega) = i& \frac{\omega_{i,0} d_{\text{eff}}}{n_i c} \mathcal{F} \left\{E_p(t) E_s^*(t) + E_{2i}(t) E_i^*(t)\right\} \nonumber \\
    -& i k_i(\omega)E_i(\omega) \label{eq:time-i} \\
    d_{z}E_{2i}(\omega) = i& \frac{\omega_{2i,0} d_{\text{eff}}}{2n_{2i} c} \mathcal{F} \{E_i^2(t)\} - i k_{2i}(\omega)E_{2i}(\omega) \label{eq:time-2i}
\end{align}
\noindent where  $E_j (\omega)=A_j (\omega)e^{i k_j (\omega_{j,0})z}$. This model allowed us to capture all non-negligible propagation effects consistent with a collinear geometry. Nonlinear polarization terms beyond quadratic order were not included.

